\documentclass[12pt]{iopart}
\usepackage{graphicx}
\usepackage{amssymb}

\newcommand{\qudit}[1]{\left\vert #1 \right\rangle}
\newcommand{\rqudit}[1]{\left\langle #1 \right\vert}

\newcommand{\E}{\mathbb{E}}

\newcommand{\N}{\mathbb{N}}
\newcommand{\A}{\mathbb{A}}
\newcommand{\R}{\mathbb{R}}

\newtheorem{thm}{Theorem}[section]
\newtheorem{cor}[thm]{Corollary}
\newtheorem{lem}[thm]{Lemma}
\newtheorem{prop}[thm]{Proposition}
\newtheorem{defn}[thm]{Definition}
\newtheorem{alg}[thm]{Algorithm}
\newtheorem{rem}[thm]{Remark}
\begin{document}

\title{Quantum sign permutation polytopes}
\author{Colin Wilmott, Hermann Kampermann and Dagmar Bru\ss}

\address{Institut f\"ur Theortische Physik III,  Heinrich-Heine-Universit\"at, D\"usseldorf, Germany}
\ead{wilmott@thphy.uni-duesseldorf.de}
\date{Received: date / Revised version: date}

%

\begin{abstract}
Convex polytopes are convex hulls of point sets in the
$n$-dimensional space $\E^n$ that generalize 2-dimensional convex
polygons and 3-dimensional convex polyhedra. We concentrate on the
class of $n$-dimensional polytopes  in $\E^n$ called sign
permutation polytopes. We characterize sign permutation polytopes
before relating their construction to constructions  over the
space of quantum density matrices. Finally, we consider the
problem of state identification  and show how sign permutation
polytopes may be useful in addressing issues of robustness.
\end{abstract}

%
\section{Introduction}
\label{intro}

From the Platonic solids to $n$-cubes, $n$-cross-polytopes and
regular $n$-simplices, convex polytopes as geometric objects have
been a source of study since antiquity. Indeed, `the Universe
cannot be read until we have learned the language in which it is
written. It is written in mathematical language, and the letters
are triangles, circles and other geometrical figures' - Galileo. A
celebrated feature of these convex geometrical figures has been
the symmetry they possess. In micro-biology where objects use the
efficiency of symmetry to propagate themselves to the study of
genetics where symmetry communicates genetic information, symmetry
has been a fundamental concept of science and has provided an
understanding for many of Nature's phenomena.  But perhaps the
most important class  of symmetric polytopes is  the class of
regular convex polytopes in Euclidean spaces.

In this paper, we concern ourselves with the class of
$n$-dimensional convex polytopes  in $\E^n$ called sign
permutation polytopes. We show that regular convex polytopes in
Euclidean space can be derived from such polytopes, and we show
that their characterization can be admitted by the partial order
of weak majorization. The outline of this paper is as follows.
Section \ref{sec:1} provides an introduction to material from
discrete geometry which serves as a basis for our study. Section
\ref{rado} introduces a theorem of Rado (Rado (1952)) and
motivates  the issue of $n$-dimensional polytopes of degree $n$ in
the $n$-dimensional
 space $\E^n$ of section \ref{construction}.  Section
\ref{quantumpolytope} relates the construction of sign permutation
polytopes to the space of density matrices in quantum mechanics,
before finally illustrating how sign permutation polytopes may be
useful in addressing the problem of robustness of state
identification.

\section{Preliminaries}
\label{sec:1} \subsection{Convex sets \& combinations}
\label{subsec:1} Let $n \in \N$ and denote by $\E^n$ the
$n$-dimensional  Euclidean space, with origin $o$, scalar product
$\langle \cdotp,\cdotp\rangle$ and induced norm $\| \cdot\|$. Let
${a}  = (\alpha_1,\alpha_2,\dots,\alpha_n) \in \E^n$ be an
$n$-tuple of real numbers. We say ${a} \in \E^n$ is a \emph{linear
combination} of  ${a}_1, \dots, {a}_m \in \E^n$ if ${a} =
\sum_{i=1}^m{\lambda_i{a}_i}$ for suitable scalars $\lambda_i \in
\R$. If such scalars $\lambda_i \in \R$, $i = 1,\dots,m$, exist
with $\sum_{i = 1}^m{\lambda_i} = 1$ then we say $a$ is an
\emph{affine combination} of ${a}_1, \dots, {a}_m \in \E^n$. The
set of points
 ${a}_1, \dots, {a}_m \in \E^n$ is said to be affinely independent if a linear combination  $\sum_{i =
1}^m{\lambda_i{a}_i}$ with $\sum_{i=1}^m{\lambda_i} = 0$ can only
have the value $o$ when $\lambda_i = 0$ for  $i = 1,\dots,m$.
 For $A \subset \E^n$, the set of all affine combinations of
points of $A$ is called the affine hull of $A$ and is denoted by
aff$A$. If ${a} = \sum_{i = 1}^m{\lambda_i{a}_i}$ for non-negative
scalars $\lambda_i \in \R$ and $\sum_{i = 1}^m{\lambda_i} = 1$
then we say $a \in \E^n$ is a \emph{convex combination} of ${a}_1,
\dots, {a}_m \in \E^n$. For $A \subset \E^n$, the set of all
convex combinations of any finitely many points of $A$ is called
the convex hull of $A$ and is denoted by conv$A$. By $H_{a,
\beta}$, we mean the hyperplane given by the linear equation of
the form $\langle a,x\rangle = \beta$. Consequently, the
hyperplane  $H_{a, \beta}$ can be expressed as the set $\{x \in
\E^n \vert \langle a,x\rangle = \beta\}$ where $ a \in
\E^n\backslash\{o\}$ and $\beta \in \R$. A closed half-space of
$\E^n$ is given by the set $\{x \in \E^n \vert \langle a,x\rangle
\geq \beta\}$, $ a \in \E^n\backslash\{o\}$ and $\beta \in \R$,
where its boundary is the hyperplane $\{x \in \E^n \vert \langle
a,x\rangle = \beta\}$, $ a \in \E^n\backslash\{o\}$ and $\beta \in
\R$. For convex sets  $A, B \subset \E^n$ satisfying ${A} \cap {
B} = \emptyset$, the Hahn-Banach separation theorem (Bishop  and
Bridges (1985)) states that there exists a unit vector $u \in
\E^n$ and a scalar $\beta \in \R$ such that for all $a \in { A}$
we have $\langle u,a\rangle \geq \beta$, while for all $a \in {B}$
we have $\langle u,a\rangle \leq \beta$.

 \subsection{The symmetric group \&  majorization} \label{subsec:2}
Let $S_n$  denote the symmetric group of degree $n$ defined as the
group  of all permutations  of the integers $1,2,\dots,n$. Denote
by $\pi({{a}}) = (\alpha_{\pi(1)}$ $,\dots, \alpha_{\pi(n)})$,
the permutation of $ a \in \E^n$ by $\pi \in S_n$. Let ${a}^* =
(\alpha_1^*,\dots,\alpha_n^*) \in \E^n$ be the point whose
components are those of $a$ arranged in non-increasing order of
magnitude; $\alpha_1^*\geq\dots \geq \alpha_n^*$ and $\alpha^*_k =
\alpha_{\pi(k)},$ for $k=1,\dots,n$  and   $\pi \in S_n$. Let
${b}^* = (\beta_1^*,\dots,\beta_n^*) \in \E^n$ be defined
analogously. If the relations
\begin{eqnarray}\label{rels}
\sum_{k\leq l}{}\alpha_k^* \leq \sum_{k\leq l}^{}{}\beta_k^*
\end{eqnarray}
hold, for $1 \leq l \leq n$, with equality for $l=n$,  we say $a$
is \emph{majorized} by $b$ and  write ${a} \prec {b}$ (Marshall
and Olkin  (1979)). For  ${{a}}, {b} \in \E^n$, we have it that
\begin{eqnarray}\label{specineq} { a} + { b} \prec {a}^*
+ { b}^*\end{eqnarray} since there exists a permutation $\pi \in
S_n$ such that $ {\alpha}_{\pi(1)}+ {\beta}_{\pi(1)} \geq
{\alpha}_{\pi(2)}+{\beta}_{\pi(2)} \geq \dots \geq
{\alpha}_{\pi(n)}+{\beta}_{\pi(n)}$ and $ \sum^{}_{k\leq
l}({\alpha}_{\pi(k)}+{\beta}_{\pi(k)}) \leq \sum^{}_{k \leq
l}({\alpha}^*_{k}+{\beta}^*_{k}),$ for $1 \leq l\leq n$ with
equality  for $l=n$. Consider the set of points ${a}_1, {a}_2,
\dots, { a}_m \in \E^n$; ${a}_i = (\alpha_{i1}, \alpha_{i2},
\dots, \alpha_{in})$,  $i = 1, 2, \dots, m$. Following from
($\ref{specineq}$), we have  it that $\label{ssp2} {a}_1 + {a}_2 +
\dots + {a}_m \prec {a}^*_1 + {a}_2^* + \dots + {a}_m^*. $
Finally, if ${a}_i \prec {b}$,  $i = 1,\dots,m,$ then $\sum_{i
\leq m}{\lambda_i}{a}_i  \prec {b}$ for non-negative scalars
$\lambda_i$ and $\sum_{i \leq  m}^{}{\lambda_i} = 1$.


\subsection{Convex polytopes} \label{subsec:3}

Convex polytopes are non-empty compact convex subsets of Euclidean
space described by a finite point set. Although the concept of
convexity is elementary,  polytopes possessing this property yield
geometric bodies with  rich structure. Two-dimensional convex
polytopes are called convex polygons. The equilateral triangle
constructed by Euclid in his first proposition is such an example.
Three-dimensional convex polytopes are called convex polyhedrons
and examples include the Platonic solids (Coxeter (1969)).
Interestingly, while  the polyhedron has eluded definition over
the centuries, the theory of convex polyhedrons and polytopes is
well understood and contributes to practically significant areas
of combinatorial optimization and computational geometry. To see
this, consider the permutahedron defined by taking the convex hull
of all vectors that are obtained by permuting the coordinates of
the vector $(1,2,\dots,n)$. A feature of the permutahedron is that
the vertices are identified with the permutations of the symmetric
group of degree $n$ in such a way that two vertices are connected
if and only if the corresponding permutations differ by an
adjacent transposition (Gaiha and Gupta (1977)).

A convex polytope is the convex hull of a non-empty finite set $A
= \{a_1, a_2, \dots, a_m\} \subset \E^n$. In particular, the
\emph{$\cal V$-polytope} is the set of points describing such a
polytope  in terms of its vertices and is given as
\begin{eqnarray}\left\{ \sum_{i = 1}^m{\lambda_i{
a}_i} \ \vert  \ \lambda_i\geq0,\  \sum_{i = 1}^m{\lambda_i} =
1\right\}.\end{eqnarray} On the other hand, a convex polytope may
be  described as the bounded solutions set of a finite system of
half-spaces in $\E^n$. In this instance, we say such a polytope is
a \emph{$\cal H$-polytope} and is described as
\begin{eqnarray}
 \{ x \in \E^n \ \vert \ \langle a_i,x \rangle \leq
\beta_i\  {\rm for} \ 1\leq i \leq m\},
\end{eqnarray}
for  $a_i \in \E^n$ and $\beta_i \in \R$. A basic result of convex
polytopes maintains that we may regard a convex polytope as a
bounded solution set of  finitely many closed half-spaces in
$\E^n$ or as the convex hull of a non-empty finite set $A \subset
\E^n$. Remarkably, while the descriptions of a convex polytope as
that of a $\cal H$-polytope or as a $\cal V$-polytope are
equivalent, the computational complexity associated with
describing each differs in the extreme. For instance, the
$n$-dimensional \emph{cross-polytope} as a $\cal V$-polytope is
defined as the convex hull of the set of $2n$ points
\begin{eqnarray}
\gamma_n = {\rm conv}\{\pi(\pm1, 0, \dots, 0), \pi \in S_n\},
\end{eqnarray}
whereas, as a $\cal H$-polytope, the $n$-cross-polytope is
described by $2^n$ half-spaces
\begin{eqnarray}
\gamma_n = \{x \in \E^n \ \vert \ \langle a_,x \rangle \leq 1\},
\end{eqnarray}
where $a$ runs through all vectors in $\{-1, 1\}^n$. The surface
of octahedron, a 3-cross-polytope,  consists of six 0-dimensional
faces called vertices, twelve 1-dimensional faces called edges and
eight 2-dimensional faces called facets. Generally, the faces of
an $n$-dimensional convex polytope have dimensions $-1, 0,1,
\dots,n$ with $-1$ denoting the dimension of the empty set with
the face of dimension $j$ being called a $j$-face. Finally, for $A
\subset \E^n$,  a convex polytope
 is called a $k$-polytope if dim ${\rm conv}\ A =
k$. This implies that some $(k+1)$-subset of $A$ is affinely
independent but  no such $(k+2)$-subset is affinely independent
(Ziegler (1995)).

\section{A note on Rado's theorem}
\label{rado}

Rado (1952) characterized the convex hull of the set of all
permutations of any real $n$-tuple in terms of the
Hardy-Littlewood-P\'{o}lya partial order relation (see Hardy \etal
(1934), p.45) for real $n$-tuples. We now state Rado's theorem and
discuss the nature  of the convex set established by the theorem.

\begin{thm}(Rado (1952)) Let  ${a}  \in \E^n$ and  ${\mathcal{V}}_{a}$ be the convex hull of
the set of permutations  of $a$.  Then ${x} \in {\mathcal{V}}_{a}$
if and only if ${x} \prec {a}$.
 \end{thm}

 The polytope of  Rado (1952)  is the $(n-1)$-dimensional permutahedron
of degree $n$    embedded in the $n$-dimensional space $\E^n$.
Since the vertices  are  obtained by permuting the integers
$(1,2,\dots,n)$, this permutahedron of degree $n$ lies entirely in
an $(n-1)$-dimensional hyperplane consisting of all points whose
coordinates sum to the integer  $1 + 2 + \dots + n = n(n+1)/2$.
For the application in quantum state identification, we concern
ourselves with the construction of $n$-dimensional polytopes of
degree $n$ in the $n$-dimensional space $\E^n$  that exploit the
key feature of Rado's theorem. We now show that a necessary and
sufficient condition for such  polytopes   requires  a  set of
$(n-1)$-dimensional bounding facets  to contain  points  whose
coordinate sums differ.

\begin{prop}\label{thm1} Let  $a,  b \in \E^n$. Let ${\mathcal{V}}_{a}$ and ${\mathcal{V}}_{b}$ denote the convex hull of
the set of permutations  of $a$ and $b$, respectively.  Then
${\mathcal{V}}_{a} \cap {\mathcal{V}}_{b} = \emptyset$ if and only
if $\sum_{k\leq n}{\alpha_k} \ne \sum_{k\leq n}{\beta_k}$.
\end{prop}

To establish proposition \ref{thm1}, we make use of the following
results. Let $A = {\rm conv}\{a_1,\dots,a_m\}\subset \E^n$ be a
non-empty convex set. By the relative interior of $A$, ${\rm
relint}{A}$,  we mean those points $a \in \E^n$ for which $a =
\sum_{i=1}^{m}{\lambda_ia_i}$ with $\lambda_i>0$, for $i =
1,\dots,k$, and   $\sum_{i=1}^{m}{\lambda_i = 1}$.

\begin{lem}\label{thm3} (Gr\"unbaum (2003))
 Let ${A}$, ${B} \subset \E^n$ be nonempty convex sets. Then ${A}$ and  ${B}$ can be properly separated if and only if
 \begin{eqnarray} {\rm relint}{A} \cap {\rm relint}{B} = \emptyset. \end{eqnarray}
 \end{lem}

\begin{lem} \label{lemma3.3} If  $a,  b \in \E^n$ such that  $\sum_{k\leq n}{\alpha_k}
= \sum_{k\leq n}{\beta_k}$, then ${\mathcal{V}}_{a}$ and
${\mathcal{V}}_{b}$ can not be properly separated by a hyperplane
$H_{{u},\alpha}$,
\begin{eqnarray}
H_{{u},\alpha} = \left\{ {z} \in \E^{n}  \ \vert \  \langle { z},{
u} \rangle = \alpha \right\},
 \end{eqnarray}
for ${u} \in \E^{n} \setminus \{o\}$, and $\alpha \in \R$.
\end{lem}

\noindent{\bf{Proof.}} See Appendix A.  \noindent{\bf{Proof of
proposition \ref{thm1}.}}


Let us suppose that  $\sum_{k\leq n}{\alpha_k} =
\sum_{k\leq n}{\beta_k}$.  
By lemma \ref{lemma3.3}, the convex hulls ${\mathcal{V}}_{a}$ and
${\mathcal{V}}_{b}$ can not be properly separated. Consequently,
by lemma \ref{thm3}, we have  ${\mathcal{V}}_{a} \cap
{\mathcal{V}}_{b} \ne \emptyset$ and the implication follows.   On
the other hand,  suppose ${\mathcal{V}}_{a} \cap {\mathcal{V}}_{b}
\ne \emptyset$. Hence there exists $ { x} \in {\mathcal{V}}_{ a}
\cap {\mathcal{V}}_{ b}$, and therefore  numbers $\lambda_i\geq 0,
\mu_i\geq 0$ satisfying   $\sum_{i\leq m}{\lambda_i}= \sum_{i\leq
m'}{\mu_i} = 1$ such that ${ x} = \sum_{i\leq m}{\lambda_i}{ a}_i$
and ${ x} = \sum_{i\leq m'}{\mu_i}{ b}_i$. For $l \leq n$ and $x =
(x_1, x_2, \dots, x_n) \in \E^n$, we have it that
\begin{eqnarray}
 \sum_{k\leq l}{x_k} = \sum_{k\leq l}\sum_{i\leq m}{}{\lambda_i\alpha_{ik}} = \sum_{i\leq m}{\lambda_i}\sum_{k\leq l}{\alpha_{ik}}
\leq \sum_{i\leq m}{\lambda_i}\sum_{k\leq l}{\alpha^*_{k}} =
\sum_{k\leq l}{\alpha^*_{k}},
\end{eqnarray}
with equality for $l = n$. Similarly, for $l \leq n$, we have it
that
\begin{eqnarray}
 \sum_{k\leq l}{x_k} = \sum_{k\leq l}\sum_{i\leq m'}{}{\mu_i\beta_{ik}} = \sum_{i\leq m'}{\mu_i}\sum_{k\leq l}{\beta_{ik}}
\leq \sum_{i\leq m'}{\mu_i}\sum_{k\leq l}{\beta^*_{k}} =
\sum_{k\leq l}{\beta^*_{k}},
\end{eqnarray}
with equality for $l = n$. The implication follows, and this
completes the proof. \hfill$\Box$

While proposition \ref{thm1} illustrates a necessary and
sufficient criterion to construct an $n$-dimensional polytope if
given an initial $(n-1)$-dimensional permutahedron, it suggests a
generalization of Rado's theorem is not readily available since
the inclusion of an arbitrary point
 does not preserve the majorization criterion. For example, consider $a,
b \in  \E^n$ and further consider the polytope described by taking
the   convex hull of   ${\mathcal{V}}_{a}$ and
${\mathcal{V}}_{b}$. Let us denote such a polytope by  ${\rm
conv}\{{\mathcal{V}}_{ a}, {\mathcal{V}}_{ b}\}$. For  some  $x
\in {\mathcal{V}}_{ a}$ and $y \in {\mathcal{V}}_{ b}$, the
question of whether a  point $z \in \E^n$ is contained in the
polytope ${\rm conv}\{{\mathcal{V}}_{ a}, {\mathcal{V}}_{ b}\}$
can be verified by showing that there exists a suitable $\lambda
\in \R$, $0\leq \lambda \leq 1$, such that $z = \lambda x +
(1-\lambda)y$. Consequently,  we can demonstrate that $z \prec
\lambda a + (1-\lambda)b$, and, hence, $z \in {\rm conv}\{{\cal
V}_a, {\cal{V}}_b\}$ but only at the insistence of a suitably
chosen  scalar $\lambda \in \R$.

\begin{figure}\label{werty}
\begin{picture}(100,145)(0,0)
\put(50,-410){\includegraphics[width=16cm]{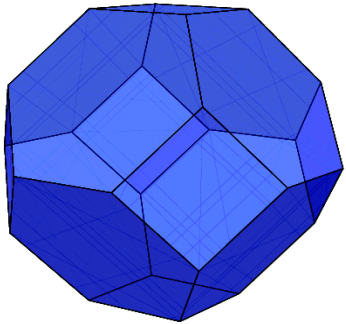}}
\end{picture}
\caption{Truncated octahedron: sign permutation polytope with
basis vector $a = (0,1,2)$ centered at the origin. The truncated
octahedron is an Archimedean solid consisting of 8 regular
hexagonal  facets, 6 square facets, 36 edges and 24 vertices.}
\label{figures3}
\end{figure}

\section{Sign permutation polytopes}
\label{construction}

Let us  instead consider the task of constructing a polytope whose
vertices are not arbitrary $n$-tuples in $\E^n$.  We adapt the
theorem of Rado (1952) to construct  an $n$-dimensional polytope
of degree $n$ in $\E^n$ with vertices given by sign permutation
points of $\E^n$ and an interior that admits a  characterization
with respect to the partial order of weak majorization.

Consider the convex polytope whose vertices consist of the set of
sign permutations on the integers $(1,2,\dots,n)$ defined as the
set of all permutations of $(\pm 1,\pm 2,\dots,\pm n)$. We call
such a polytope consisting of $2^nn!$ vertices a  sign permutation
polytope, see figure \ref{figures3}. The sign permutation polytope
is $n$-dimensional convex  polytope of degree $n$ in the
$n$-dimensional space $\E^n$. More generally, denote by
${\cal{V}}_{\pm \pi({a})}$  the sign permutation polytope given as
the convex hull of componentwise  sign permutation changes of ${a}
\in \E^n$;
\begin{eqnarray}
{\cal{V}}_{\pm \pi({a})}= {\rm conv}\{(\pm \alpha_{\pi(1)}, \pm
\alpha_{\pi(2)}, \dots, \pm \alpha_{\pi(n)}), \pi \in S_n\}.
\end{eqnarray}
Let    ${a}^* \in \E^n$ denote again that vector whose components
are those of $a$ arranged in non-increasing order of magnitude.
Let $b\in \E^n$ be defined analogously.  If the relations
\begin{eqnarray}\label{rels} \sum_{k \leq l}{}\alpha_k^* \leq
\sum_{k\leq l}{}\beta_k^*
\end{eqnarray}
hold, for $1 \leq l \leq n-1$,  and
\begin{eqnarray}\label{rels2}\sum_{k \leq n}\alpha_k^* <
\sum_{k \leq n}{}\beta_k^*, \end{eqnarray}   we then say $a$ is
\emph{weakly majorized} by $b$, and write  $a  \prec_{w} b$. Let
$\E^n_+$ denote  the set $\E^n_+ = \{(\alpha_1,\dots,\alpha_n) \ |
\ \alpha_i \geq 0, \ i = 1,\dots,n\}$.

\begin{cor}\label{thm5}  Let ${a}, {x} \in \E_{{+}}^{n}$. Then  ${x} \in {\cal{V}}_{\pm \pi({a})}$  if
and only if ${x} \prec_{w} {a}$.
\end{cor}

\noindent{\bf{Proof.}} If ${ x} \in {\mathcal{V}}_{\pm\pi({a})}$,
then there are numbers $\lambda_i$ and vectors ${a}_i$, $i = 1, 2,
\dots, m$, for $m \leq 2^n n!$ such that  ${a}_i \prec_{w}  a$,
$\lambda_i \geq 0$, $\sum_{i\leq m }^{}{\lambda_i} = 1$ and $ x =
\sum_{i\leq m}^{}{\lambda_i{a}_i}$. Then since ${a}_i \prec_{w}
a$, we have it that  ${x} = \sum_{i\leq m}^{}{\lambda_i{a}_i}
\prec_{w} \sum_{i\leq m}^{}{}\lambda_i{a} = a$ for $\lambda_i \geq
0$, $i=1,2,\dots,m$, and  $\sum_{i\leq m }^{}{\lambda_i} = 1$.
On the other hand,  suppose ${ x} \prec_{w} { a}$ while ${ x}
\notin
{\mathcal{V}}_{\pm\pi({a})}$. Using the Hahn-Banach separation argument of Rado (1952), the reverse implication is readily established,
and the result follows. \hfill $\Box$ 

We make note of the following relationship between the convex
hulls of ${\cal V}_{\pm{\pi(a)}}$ and ${\cal V}_{\pm{\pi(x)}}$.
\begin{cor}\label{corr4} Let ${a}, {x} \in \E_+^n$. Let ${\cal V}_{\pm{\pi(a)}}$ and ${\cal V}_{\pm{\pi(x)}}$ denote the convex
hull of  sign  permutations of ${a}, {x} \in \E_+^n$,
respectively. Suppose  ${x} \prec_{w} {a}.$ Then ${\cal
V}_{\pm{\pi(x)}} \subseteq {\cal V}_{\pm{\pi(a)}}$.
\end{cor}
\noindent{\bf{Proof.}} Given  ${a}, {x} \in \E_+^n$ with ${x}
\prec_{w} {a}$, we have it that $\pm\pi({x}) \prec_{w} {x}$ for
$\pi \in S_n$. Noting that $\prec_{w}$ is a partial order then
transitivity of $\prec_{w}$  establishes $\pm\pi({x}) \prec_{w}
\pm\pi({a})$, and the result follows.  \hfill$\Box$

\begin{rem}
We note that Mirsky (1959) has constructed a polytope similar to
the polytope of corollary \ref{thm5}  restricted to the positive
orthant of $n$-dimensional Euclidean space.   Markus (1964)
characterized the boundary of the sign permutation polytope.
\end{rem}

A regular polytope  generalizes  the Platonic solids to arbitrary
dimensions.  For the $n$-dimensional spaces $\E^n$ with $n \geq
5$, there are only three regular convex polytopes (Coxeter
(1948)). These are the $n$-dimensional regular simplex, the
$n$-dimensional cross-polytope and the $n$-dimensional cube. For
${a} = (\alpha,0,\dots,0) \in \E^n$, the sign permutation polytope
${\cal{V}}_{\pm \pi({a})}$ degenerates to  the class of
cross-polytopes $\beta_n$. The cross-polytope $\beta_n$ can be
described in terms of the  $2n$ vertex set
$\{\pi(\pm\alpha,0,\dots,0), \ \pi \in S_n\}$, or, equally, in
terms of the  $2^n$ half spaces $\gamma_n = \{x \in \E^n \ \vert \
\langle a_,x \rangle \leq 1\},$
where $a$ runs through all vectors in $\{-1, 1\}^n$. 
 For ${a} =
(\alpha,\alpha,\dots,\alpha) \in \E^n$, the sign permutation
polytope ${\cal{V}}_{\pm \pi({a})}$ coincides with the $n$-cube
$\gamma_n$ which  can be evaluated either in terms of the $2n$
half-spaces $\{ x \in \E^n \ \vert \ -\alpha \leq x_k \leq \alpha,
\ {\rm for}\ 1\leq k \leq n\}$, or, equivalently,  with respect to
the $2^n$ vertex set $\{(\pm\alpha,\pm\alpha,\dots,\pm\alpha)\}. $
More generally, if we consider the $n$-tuple of real coefficients
$a = (\alpha_1,\dots,\alpha_m,0_1,\dots,0_{n-m}) \in \E^n$
consisting of $k\leq n$ distinct coefficients of multiplicity
$m_i$, $1\leq i\leq k$, then the number of distinct sign
permutations, i.e. vertices, in the set
$\{\pi(\pm\alpha_1,\dots,\pm\alpha_m,0_1,\dots,0_{n-m}), \ \pi \in
S_n\}$
   is
\begin{eqnarray}
\frac{2^nn!}{2^{n-m} \ m_1!\dots m_{k-1}!(n-m)!}.
\end{eqnarray}
Consequently, we have it that the sign permutation polytope
${\cal{V}}_{\pm \pi({a})}$ given by
\begin{eqnarray}
{\rm conv}\{\pi(\pm\alpha_1,\dots,\pm\alpha_m,0_1,\dots,0_{n-m}),
\ \pi \in S_n\},
\end{eqnarray}
  has a vertex set of order
${2^mn!}/{({m_1!\dots m_{k-1}!(n-m)!})}.$ We see again that for
$m=1$, the sign permutation polytope ${\cal{V}}_{\pm \pi({a})}$,
$a = (\alpha_1,\dots,\alpha_m,0_1,\dots,0_{n-m})$ $\in \E^n$,
$m\leq n$, describes the $n$-dimensional cross-polytope $\beta_n$,
and for $m=n$ and $k = n$, the sign permutation polytope coincides
with the $n$-dimensional cube $\gamma_n$. For ${a} =
(\alpha,0,\dots,0) \in \E^{n+1}$, we note that the $n$-dimensional
regular simplex $\alpha_n$ is a permutation polytope of degree
$n+1$ defined as the convex hull of all permutations of the vector
${a} \in \E^{n+1}$.

Finally, for what follows, we derive the volume and insphere
radius for the $n$-cross-polytope.

\begin{lem}\label{lemma1}
Let $\alpha \geq 0$. The volume of the $n$-dimensional
cross-polytope ${\rm conv}\{\pi(\pm \alpha, 0, \dots, 0)$, $\pi
\in S_n\}$ of edge length $\sqrt{2}\alpha$ is
${(2\alpha)^{n}}/{(n)!}$.
\end{lem}
The cross-polytope $\beta_n$ is a composition of $2^n$ convex
regions with each region corresponding to an orthant of $\beta_n$.
The convex region determined by the positive orthant of $\beta_n$
is an $n$-dimensional cone of height $\alpha$. In $\E^n$, the
volume of the cone of height $h$ over an $(n-1)$-dimensional base
of volume $B$ is given as $Bh/n$. By induction, the volume of the
cone in the positive orthant of $\beta_n$ is $\alpha^n/n!$.
Consequently, the volume of $\beta_n$ is ${(2\alpha)^{n}}/{n!}$.

Next we consider  the inscribed sphere of the cross-polytope.

\begin{lem}\label{lemma2}
The  insphere of the  cross-polytope conv$\{\pi(\pm \alpha, 0,
\dots, 0), \pi \in S_n\}$, $\alpha \geq 0$, of side length
$\sqrt{2}\alpha$ has radius $r_{\rm in} = \alpha/\sqrt{n}.$
\end{lem}

\noindent{\bf{Proof.}} We maximize the $n$-dimensional sphere
$\sum_{i=1}^{n}{x_i}^2 = r^2$  subject to the constraint
$\sum_{i=1}^{n}{x_i} = \alpha$. The constraint
$\sum_{i=1}^{n}{x_i} = \alpha$ represents the coordinate sum of
the points on the facet of the cross-polytope in the positive
orthant of $\E^n$.
Let $L(x_1,x_2,\dots,x_n) = \sum_{i=1}^{n}{x_i}^2 -
\lambda\left(\sum_{i=1}^{n}{x_i} - \alpha\right).$ By the method
of Lagrange multipliers, we have it that $\nabla
L(x_1,x_2,\dots,x_n)$ $= \{2x_i - \lambda, i = 1,\dots,n$\} with
$\sum_{i=1}^{n}{x_i} = \alpha$. Solving the system of equations
yields $x_i = \alpha/n$ for $i=1,\dots,n$, and the result follows.
\hfill$\Box$

%

\section{Quantum sign permutation polytopes}
\label{quantumpolytope}

A fundamental topic of quantum information theory has been the
characterization of entanglement amongst states of a quantum
system.  A natural question concerns the  partitioning of the set
of all states of some finite dimensional composite quantum system
according to   entanglement type. It has been shown that all
bipartite entangled pure states are asymptotically equivalent, up
to local operations and classical communication, to the
Einstein-Podolsky-Rosen state (Bennett \etal (1996)). For systems
of three or more parties, there are several equivalence classes of
different  entanglement types  (Vidal (2000)). We note the
particular instance of three qubit systems in which   D\"ur \etal
(2000) has shown that  there are two equivalence classes
possessing genuine
tripartite entanglement. 

In this article, we  focus on the construction of sign
permutations polytopes within the space of quantum states, paying
attention to three-qubit systems. We demonstrate the construction
of quantum sign permutation polytopes for both pure and mixed
quantum states, and we utilize the weak majorization feature
admitted by sign permutation polytopes to classify a certain
subset of quantum states.  The construction of sign permutation
polytopes from section \ref{construction} can be translated to the
quantum world as follows.

Let $d \in \N$ and let $\cal{H}$ denote the $d$-dimensional
complex Hilbert space. Let
 ${\cal{M}}_d$ denote the set of  states that coincide with the set of $d\times d$ complex Hermitian matrices with non-negative eigenvalues and unit trace
 which act on the Hilbert space $\cal{H}$;
\begin{eqnarray}
 {\cal M}_d = \{ \varrho \ \vert \  \varrho = \varrho^\dagger; \varrho \geq 0; {\rm Tr}\varrho = 1\}.
\end{eqnarray}
${\cal M}_d$ is a $(d^2-1)$-real-dimensional convex set and the
vertices of this set form  a $2(d-1)$-dimensional subspace of the
$(d^2-2)$-dimensional boundary $\partial{{\cal M}_d}$. The set of
density matrices of ${\cal M}_d$ consists  of pure and mixed
states and
 the pure states correspond to the extreme points of this set. For $\varrho
\in {\cal{M}}_d$,  $\varrho$ is a pure state if and only if
$\varrho^2 = \varrho$ with $\varrho$ written as a rank one
projector onto $\cal{H}$. Otherwise, $\varrho$ is a mixed state,
and we write $\varrho$ as a convex combination of pure states.
Noting that any general $n$-dimensional affine space $\A$ is
isomorphic the $n$-dimensional affine space $\E^n$ (Br\o{}ndsted
(1983)), we put $n = d^2-1$, and introduce a correspondence
between a state $\varrho_{a} \in {\cal M}_d$ and a point  ${{a}}
\in \ \E^{d^2-1}$ by \clearpage

\begin{eqnarray}
 \hskip12.5em{\cal{M}}_d &\rightarrow& \ \ \ {\E^{d^2-1}}\\ \varrho_a = \left(%
\begin{array}{ccc}
   \alpha_1 & \alpha_2 + \alpha_3\iota & \dots \\
  \alpha_2 - \alpha_3\iota &  \alpha_4  & \dots \\
   &    &\ddots\\
\end{array}%
\right) &\mapsto& \left(\begin{array}{c}
   \alpha_1 \\
  \alpha_2\\
  \alpha_3  \\
\alpha_4\\ \vdots \\
   \\
\end{array}\right) = a.
\end{eqnarray}
To ensure that   $n$-dimensional polytopes  have  non-zero volume
in $\E^n$, we require that the dimension of the corresponding
affine hull be $n+1$. This fact may be demonstrated by noting that
affine independence of an $(n+1)$-family of points in $\E^n$ is
established if and only if there is  linear independence amongst
an augmentation of the $(n+1)$-family of points in $\E^{n+1}$.
\begin{lem}(Br\o{}ndsted (1983))  \label{lemma6} Let $\{{a}_i\} \in \E^n$, ${i = 0,\dots, n}$, and let $\tau({a}): \E^n \rightarrow \E^{n+1}$ be defined by $\tau({a}) = ({a},1)$.
The set  $\{{a}_i\}$, ${i = 0,\dots, n}$, is affinely independent
in $\E^n$ if and only if $\{({a}_i,1)\}$, ${i = 0,\dots, n}$, is
linearly independent in $\E^{n+1}$.
\end{lem}\noindent{\bf{Proof.}} Suppose that $\{{a}_i\}$, ${i = 0,\dots, n}$, is affinely
independent in $\E^n$, i.e., $\sum_{i=0}^{n}{\lambda_i{a}_i}$
vanishes in $\E^n$ and  $\sum_{i=0}^{n}{\lambda_i}$ vanishes in
$\R$ only when $\lambda_i = 0$,  $i = 0,\dots, n$. In particular,
we have $(\sum_{i=0}^{n}{\lambda_i{{a}}_i},
\sum_{i=0}^{n}{\lambda_i})$ $=
\sum_{i=0}^{n}{\lambda_i({{a}}_i,1)}$ $= (0_{\E^n}, 0_{\R})$ $=
0_{\E^{n+1}}$ only when $\lambda_i = 0$, $i = 0, \dots,n$.
Consequently, $\{({{ a}}_i,1)\}$, ${i = 0,\dots, n}$, is linearly
independent in $\E^{n+1}.$ \noindent Conversely, suppose that
$\{({{ a}}_i,1)\}$, ${i = 0,\dots, n}$, is linearly independent in
$\E^{n+1}$, i.e., $\sum_{i=1}^{n}{\mu_i({{a}}_i,1)}$ $=
0_{\E^{n+1}}$ only when $\mu_i = 0$, $i = 0,\dots, n$. Therefore,
$\sum_{i=0}^{n}{\mu_i{{ a}}_i} = 0_{\E^n}$ and
$\sum_{i=0}^{n}{\mu_i} = 0_{\R}$ occurs only when $\mu_i = 0$, $i
= 0,\dots, n$. Thus, $\{{{a}}_i\}$, ${i = 0,\dots,n}$, is affinely
independent in $\E^n$, and the result follows. \hfill$\Box$

\subsection{Pure state polytopes}\label{purequantumpolytope}

\begin{defn} A \emph{pure state polytope} is a polytope whose
vertex description is  given by a set of pure states, i.e.,
conv$\{\varrho_i\}$ where $\varrho_i \in {\mathcal{M}}_d $ such
that  ${\rm Tr}\varrho_i^2 = 1, \varrho_i = \varrho_i^\dagger,
\varrho_i \geq 0$ and ${\rm Tr}\varrho_i = 1.$ A \emph{pure state
sign permutation polytope} is a polytope whose vertex set is
constructed by applying the set of sign permutations on a pure
state and admitting only those sign permutations that yield a
valid pure quantum state.
\end{defn}

\begin{figure}
\begin{picture}(100,130)(0,0)
\put(145,-40){\includegraphics[width=6cm]{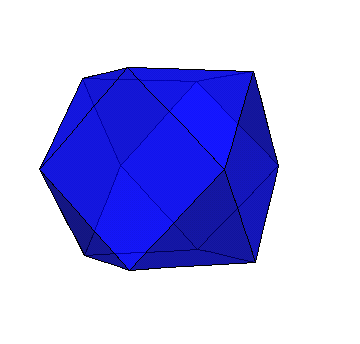}}
\end{picture}
\caption{Qubit Cuboctahedron: sign permutation polytope with basis
vector $(1/\sqrt{2},0,1/\sqrt{2})$. The cuboctahedron  consists of
8 triangular faces, six square faces, 24 edges and 12
vertices.}\label{cube}
\end{figure}

 Let $\varrho \in {\mathcal{M}}_2$ be a pure qubit state,
\begin{eqnarray} \varrho = 1/2\left(\begin{array}{cc}
  1 + P_z & P_x - iP_y \\
  P_x - iP_y & 1 - P_z \\
\end{array}\right)\nonumber\end{eqnarray}
with ${P_x}^2+{P_y}^2+{P_z}^2 = 1$ and ${P_x}, {P_y}, {P_z} \in
\R$. The parametrization of $\varrho \in {\mathcal{M}}_2$ is given
by $({P_x}, {P_y}, {P_z}) \in \R^3$. Applying the  set of sign
permutations to the  qubit parametrization $(1/\sqrt{2}, 0,
1/\sqrt{2})$ and taking only those sign permutations that return
valid pure quantum states yields the cuboctahedron of Figure \ref{cube}.\\

For the construction of a quantum sign permutation polytope from a
pure state $\qudit{\psi} \in {\cal H}$, we enumerate the set of
sign permutations on the Euclidean representative of density
matrix
$\qudit{\psi}\rqudit{\psi} \in {\cal{M}}_d$ and   
consider only those sign permutations $\pm\pi_{\rm{pure}}$ for
which $\pm\pi_{\rm{pure}}(\qudit{\psi}\rqudit{\psi})$  is again a
pure state. For three-qubit systems, the general description of a
pure state $\qudit{\psi}$ is given by
\begin{eqnarray}\label{general}
\qudit{\psi} = \lambda_0\qudit{000}+\lambda_1e^{\iota
\theta}\qudit{100}+\lambda_2\qudit{101} +
\lambda_3\qudit{110}+\lambda_4\qudit{111}
\end{eqnarray}
where $\lambda_i \geq 0, 0\leq\theta\leq\pi$ and
$\sum_i{\lambda^2_i = 1}$ (Ac\'{i}n \etal (2001)).  D\"ur {\etal}
(2000) have shown that non-trivial tripartite
entanglement of three qubit systems can 
 be described in terms of two equivalence classes. In particular, a state  in
the system of three qubits  possessing genuine tripartite
entanglement may be a Greenberger-Horne-Zeilinger (GHZ) state for
which  a typical  representative state  is given by
\begin{eqnarray} \qudit{{\rm GHZ}} = {1}/{\sqrt{2}}\left(\qudit{000}
+\qudit{111}\right). \end{eqnarray}   On the other hand, a
three-qubit state may  be in  the W class of tripartite entangled
states. A quantum state is said to be  W-type entangled if
$\lambda_4 = \theta = 0$ in (\ref{general}). A typical
representative state for those states possessing W-type
entanglement  is given by
\begin{eqnarray} \qudit{W} = {1}/{\sqrt{3}}\left(\qudit{100}
 +\qudit{010}+\qudit{001}\right).\end{eqnarray}

As an example, we  construct a quantum polytope of high
dimensionality for some fixed affine space that   possesses W-type
entanglement. For such a polytope, we considered a random
three-qubit pure W-state $\qudit{W} = $ $0.758\iota\qudit{0}$ $+
(0.809 - 0.588\iota)\qudit{2}$ $+ (0.809 + 0.588\iota)\qudit{5}$
$+0.242\qudit{7}$. Taking the set of all possible sign
permutations on the parameterized vector, we construct the
polytope vertex set by  only including those sign permutations
that correspond that valid W-type entangled pure quantum states.
W-type entanglement is ensured by showing a vanishing  3-tangle on
the set of valid quantum states. The 3-tangle, $\tau_3$, is the
measure most often utilized to distinguish between the GHZ- and W-
class of three-qubit states (Coffman \etal  (2000)). For
 pure states,  the 3-tangle is equivalent to the absolute value of a quantity
known as Cayley's hyperdeterminant (Cayley (1845)), and is known
to vanish for those states possessing W-type entanglement. In our
example, there are $2^48!/4! = 26880$ possible sign  permutations
for a general three-qubit W-type entangled state of which a total
of 5376 pure points are returned as possessing  W-type
entanglement.

\subsection{Mixed state polytopes}
\label{mixedquantumpolytope}

\begin{defn} A \emph{mixed state polytope} is a polytope whose
vertex description is given by a set of mixed states, i.e.,
conv$\{\varrho_i\}$ where $\varrho_i \in {\mathcal{M}}_d $ such
that  ${\rm Tr}\varrho_i^2 < 1, \varrho_i = \varrho_i^\dagger,
\varrho_i \geq 0$ and ${\rm Tr}\varrho_i = 1.$ A \emph{mixed state
sign permutation polytope} is a polytope whose vertex set is
constructed by applying the set of sign permutations on a mixed
state and admitting only those sign permutations that yield a
valid  quantum state.
\end{defn}

An open problem of multipartite entanglement is the identification
of the  entanglement type for a given  quantum state. Appealing to
the theory of convex sets may help resolve the issue of
multipartite entanglement identification whereby a state of
unknown entanglement type is described  as a convex combination of
states of a  fixed entanglement type. As noise has always to be
considered for an experiment, we are interested in describing a
volume (an $\epsilon$-ball) around the given  state which is still
of the same entanglement type. However, achieving a lower bound on
the allowed $\epsilon$-environment requires knowledge about the
facet set of a polytope  of a fixed entanglement type. For
$n$-dimensional convex polytopes consisting of a vertex set of
order $m$, the set of hyperplanes can be of the order $m^{\lfloor
n/2\rfloor}$ (Matousek  (2002)). In this instance, an
$\epsilon$-ball argument to identify multipartite entanglement is
rendered inefficient.

Let us now suppose that there exists  a convex decomposition of a
given state $\varrho \in {\cal M}_d$ in terms of  states of a
fixed entanglement type (Kampermann \etal (2010)). We consider the
construction of quantum sign permutation polytopes whose vertex
set is now contained in the convex hull of the given convex
decomposition. The volume of the constructed sign permutation
polytope allows conclusions about the  robustness of state
identification.  For  efficiency in high dimensions, we consider
the restriction of a general sign permutation polytope to that of
the $n$-cross-polytope.  The  quantum polytope analysis proceeds
in two parts. The initial part of the process is concerned with
the construction of the  regular $n$-cross-polytope. In a second
algorithm we  address the robustness of the  identification
process.

Algorithm \ref{algorithm1} constructs a sign permutation polytope
of  edge length $\sqrt{2}\alpha$, $\alpha \in \R$, centered on a
given state $\varrho \in {\cal{M}}_d$. The resulting polytope
describes a convex hull of states of a certain entanglement type
by ensuring that the vertices of the constructed quantum polytope
be written as a convex composition of points from the initial
convex set of states possessing a fixed entanglement type. We
assume that the convex set input of algorithm \ref{algorithm1} has
order $m>d^2-1$. This is because we wish construct polytopes with
non-zero volume centered on a mixed state in ${\mathcal{M}}_d$.

\clearpage

\begin{alg}\label{algorithm1}
{{\textbf{Quantum polytope construction}}}

\noindent\underline{\textbf{Input:}}\begin{enumerate}
    \item Given state  $\varrho \in {\cal M}_d$
    \item   Convex set of states of a fixed entanglement type, $\varrho_i \in {\cal M}_d$ for $i = 1,\dots,m$
    and $m>d^2-1$, such that $\varrho = \sum^{m}_{i=1}{}\lambda_i\varrho_i, \lambda_i\geq 0$ and $\sum^{m}_{i=1}{}\lambda_i = 1.$ (Kampermann {et al} (2010))
\end{enumerate}

\begin{figure}
\begin{picture}(10,200)(0,0)
\put(30,125){\includegraphics[width=3cm]{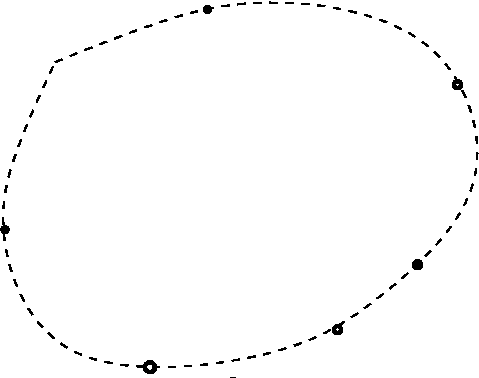}}
\put(190,123){\includegraphics[width=4cm]{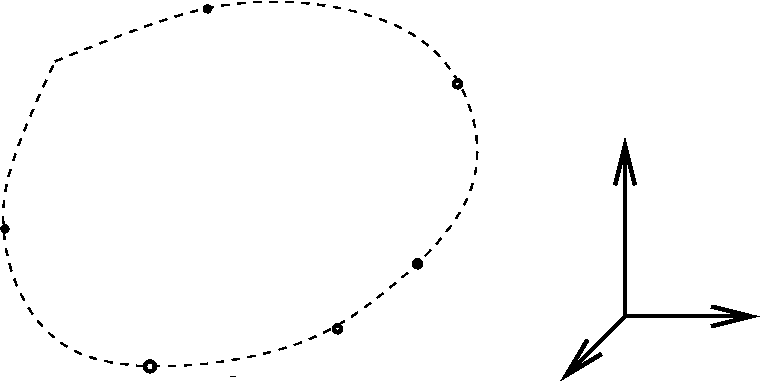}}
\put(352,63){\includegraphics[width=3cm]{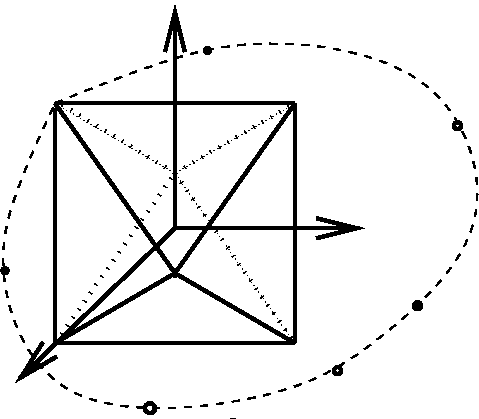}}
\put(192,7){\includegraphics[width=4cm]{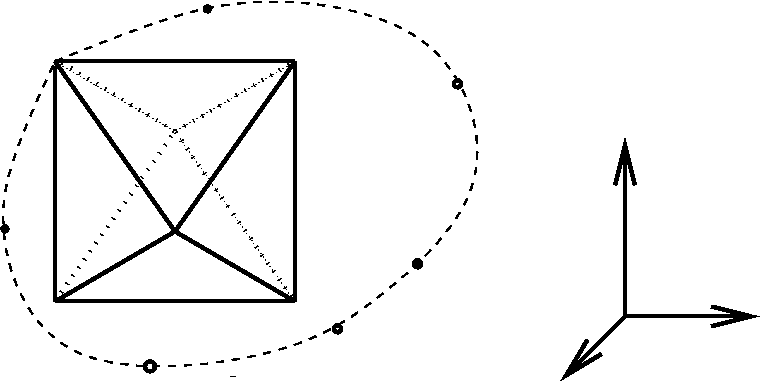}}
\put(30,15){\includegraphics[width=3cm]{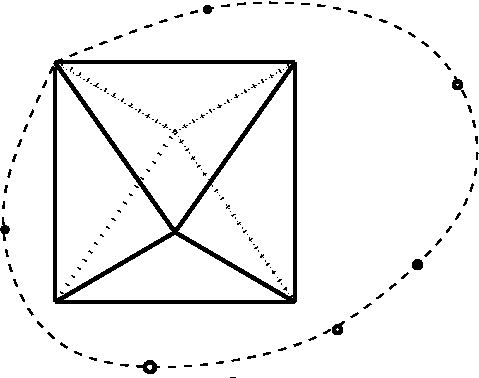}}

\put(05,180){${}_{{1.}}$}\put(65,150){${}_{\varrho}$}\put(59,151){${}_{\star}$}\put(35,190){${}_{\varrho_1}$}\put(105,190){${}_{{\cal
M}_d}$}\put(73,198){${}_{\varrho_2}$}\put(80,196){${}_{\dots}$}\put(15,155){${}_{\varrho_m}$}

\put(175,180){${}_{{2.}}$}\put(255,180){${}_{{\E}^{d^2-1}}$}\put(220,144){${}_{\tilde{\varrho}}$}
\put(215,148){${}_{\star}$}\put(195,180){${}_{\tilde{\varrho}_1}$}\put(220,188){${}_{\tilde{\varrho}_2}$}\put(230,185){${}_{\dots}$}\put(175,150){${}_{\tilde{\varrho}_m}$}

\put(335,132){${}_{{3.}}$}\put(440,125){${}_{{\E}^{d^2-1}}$}
\put(380,96.5){${}_{\star}$}\put(330,116){${}_{T_{\tilde{\varrho}}(\tilde{\varrho}_1)}$}\put(388,137){${}_{T_{\tilde{\varrho}}
(\tilde{\varrho}_2)}$}\put(400,133){${}_{\dots}$}\put(315,95){${}_{T_{\tilde{\varrho}}(\tilde{\varrho}_m)}$}

\put(175,70){${}_{{4.}}$}\put(255,70){${}_{{\E}^{d^2-1}}$}\put(225,34){${}_{\tilde{\varrho}}$}\put(216,37){${}_{\star}$}\put(190,54){${}_{\tilde{\varrho}_1}$}\put(215,67){${}_{\tilde{\varrho}_2}$}\put(225,67){${}_{\dots}$}\put(180,29){${}_{\tilde{\varrho}_m}$}

\put(05,70){${}_{{5.}}$} \put(105,80){${}_{{\cal
M}_d}$}\put(66,43){${}_{{\varrho}}$}\put(59,48.5){${}_{\star}$}\put(30,71){${}_{{\varrho}_1}$}\put(66,87){${}_{{\varrho}_2}$}
\put(74,85){${}_{\dots}$}\put(15,50){${}_{{\varrho}_m}$}

\end{picture}
\caption{Constructing a quantum $(d^2-1)$-cross-polytope:
${\mathcal{V}}_{\pm\pi(a)}$, $\pi \in S_{d^2-1}$ with basis vector
$a = (\alpha,0,\dots,0)$. 1.\hskip.5emillustrates the given state
$\varrho$ as a convex combination of states of a fixed
entanglement type.  2.\hskip.5em3.\hskip.5emand
4.\hskip.5emillustrate the construction of
${\mathcal{V}}_{\pm\pi(a)}$ centered on $\tilde{\varrho}$ while
5.\hskip.5emshows a  $(d^2-1)$-quantum-cross-polytope centered on
the given state $\varrho \in {\cal M}_d$.} \label{figures1}
\end{figure}
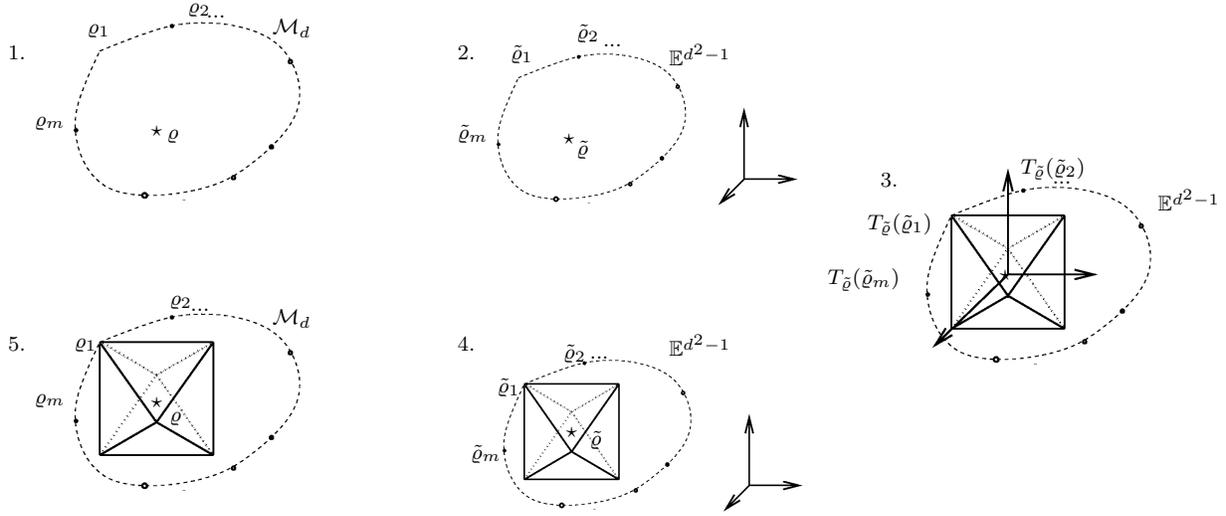

\noindent\underline{\textbf{Output:}} Quantum
$(d^2-1)$-cross-polytope of a fixed entanglement type with edge
length $\sqrt{2}\alpha$ centered at
 $\varrho$,    volume $(2\alpha)^{d^2-1}/(d^2-1)!$  and  an inscribed sphere of radius $r_{\rm in} = \alpha/\sqrt{d^2-1}$, see lemma \ref{lemma1} and lemma \ref{lemma2}.

\noindent\underline{\textbf{Procedure:}}
\begin{enumerate}
    \item Let  $\tilde{\varrho} \in \E^{d^2-1}$ denote the Euclidean vector
    associated with $\varrho \in {\cal M}_d$. Let $\tilde{\varrho_i} \in
    \E^{d^2-1}, \ i = 1,\dots, m$, be analogously defined.
    \item Let $T_{\tilde{\varrho}}(a): a - \tilde{\varrho}$,  $a \in \E^{d^2-1}$, be an affine
    transformation on $\E^{d^2-1}$. Let ${\rm conv}\{T_{\tilde{\varrho}}(\tilde{\varrho_i}), \ i =
    1,\dots,m\}$ be  the translate  of ${\rm conv}\{\tilde{\varrho_i}, \ i = 1,\dots,    m\}$ under $T_{\tilde{\varrho}}$.
    \item For suitable $\alpha \in \R$, we ensure that the set of
sign permutations of the vector $(\alpha, 0, \dots, 0) \in
\E_+^{d^2-1}$ is contained in the  convex hull of
$T_{\tilde{\varrho}}(\tilde{\varrho_i}), \ i = 1,\dots, m$.
   \item Maximise $\alpha \in \R$ such that the set of $2(d^2-1)$ points  $\{\pi(\pm \alpha, 0, \dots, 0),  \pi \in S_{d^2-1}\}$  are still contained in the convex hull of $T_{\tilde{\varrho}}(\tilde{\varrho_i}), \ i = 1,\dots,
    m$.
    \item Let $T^{-1}_{\tilde{\varrho}}(a)$ be an affine
    transformation on $\E^{d^2-1}$ given by $T^{-1}_{\tilde{\varrho}}(a): a + \tilde{\varrho}$,  $a \in \E^{d^2-1}$.
    Then
    \begin{eqnarray}
     {\rm conv}\{T^{-1}_{\tilde{\varrho}}(\{\pi(\pm \alpha, 0, \dots, 0),  \pi \in S_{d^2-1}\})\}
    \end{eqnarray}
   is the $(d^2-1)$-dimensional  cross-polytope  with edge length $\sqrt{2}\alpha$ and volume $(2\alpha)^{d^2-1}/(d^2-1)!$ centered at $\tilde{\varrho}$
    and contained in the convex hull  conv$\{\tilde{\varrho_i},  i = 1,\dots,
    m\}$.
    \end{enumerate}
\end{alg}

Step (i) of algorithm \ref{algorithm1}  establishes  a
correspondence between the set of  density operators ${\cal{M}}_d$
and  vectors in Euclidean space $\E^{d^2-1}$, see figure
\ref{figures1} (1.\hskip.5em and 2.). The quantum states $\varrho$
and $ {\varrho_i}$, for $i = 1,\dots, m,$ are mapped to their
corresponding representatives in $\E^{d^2-1}$. Step (ii) describes
the translation transformation $T_{\tilde{\varrho}}$ that acts on
the set of classical vectors $\tilde{\varrho}$ and
$\tilde{\varrho_i},$ for $i = 1,\dots, m$. Under
$T_{\tilde{\varrho}}$, the vector $\tilde{\varrho} \in \E^{d^2-1}$
is mapped to the $o \in \E^{d^2-1}$. Correspondingly, the set
$\tilde{\varrho_i}$, for  $i = 1,\dots, m$, is mapped to a
surrounding neighbourhood of the origin. Step (iii) then
constructs a sign permutation polytope with edge length
$\sqrt{2}\alpha$ centered at origin for the basis vector $a =
(\alpha, 0, \dots, 0) \in \E_+^{d^2-1}$, see figure \ref{figures1}
(3.). Step (iv) utilizes  a divide and conquer algorithm (O'
Rourke (1998)) to ensure  that the value  $\alpha \in \R$ in the
basis vector induces the polytope of maximum volume. By lemma
\ref{lemma2}, the constructed sign permutation polytope has volume
$(2\alpha)^{d^2-1}/(d^2-1)!$. Since this polytope is defined by
taking the sign permutations of the vector $(\alpha, 0, \dots, 0)
\in \E_+^{d^2-1}$,  the construction yields the regular
$(d^2-1)$-dimensional cross-polytope. Step (v) implements the
translation transformation $T^{-1}_{\tilde{\varrho}}$ on the set
of  vertices of the constructed cross-polytope. Since a translate
of a convex polytope is again  a convex polytope, a consequence of
result $x + {\rm conv}M = {\rm conv}(x + M)$ (Br\o{}ndsted
(1983)),  for $x\in \E^n$ and $M \subset \E^n$,
    it follows that
${\rm conv}\{T^{-1}_{\tilde{\varrho}}(\{\pi(\pm \alpha, 0, \dots,
0), \pi \in S_{d^2-1}\})\}$ is a regular cross-polytope centered
at $\tilde{\varrho}$ with edge length $\sqrt{2}\alpha$ and volume
$(2\alpha)^{d^2-1}/(d^2-1)!$, see figure \ref{figures1} (4.).
Since any $(d^2-1)$-dimensional affine space is isomorphic to the
Euclidean space $\E^{d^2-1}$,  the constructed polytope yields a
desired quantum polytope in ${\cal{M}}_d$ with respect to the
Hilbert-Schmidt norm, see figure \ref{figures1} (5.\hskip-0em).
\hfill$\Box$

The next stage in the process addresses the question of robustness
of multipartite entanglement identification. Algorithm
\ref{algorithm2} makes use of the weak majorization criterion to
determine the points $\varrho' \in {\cal M}_d$ in the convex hull
of a constructed sign permutation polytope. The set of points
$\varrho' \in {\cal M}_d$ can be  represented  as the  state
$\varrho \in {\cal M}_d$ plus noise in the quantum system. We make
use of the following result from \.{Z}yczkowski and Sommers
(2003).

\begin{thm}\label{thm2}(\.{Z}yczkowski and Sommers (2003))
The volume of the $(d^2-1)$-dimensional set  of states ${\cal
M}_d$ with respect to the Hilbert-Schmidt  measure is given by
\begin{eqnarray}
{\rm Vol_{HS}}({\cal M}_d) =
\sqrt{d}\frac{\pi^{d(d-1)/2}}{2^{(d-1)/2}}\frac{\Gamma(1)\dots\Gamma(d)}{\Gamma(d^2)}
\end{eqnarray}
where $\Gamma(\cdot)$ is the Euler gamma function.
\end{thm}

\begin{alg}\label{algorithm2}
{{\textbf{On the robustness of state identification}}}

\noindent\underline{\textbf{Input:}} \begin{enumerate}
    \item Cross polytope conv$\{\pi(\pm \alpha, 0, \dots, 0),  \pi \in S_{d^2-1}\}$
    of edge length $\sqrt{2}\alpha$ and volume $(2\alpha)^{d^2-1}/{d^2-1}!$ associated with a quantum polytope of a fixed  entanglement type.
 \item  Given state $\varrho \in {\cal M}_d$.
    \item An arbitrary state $\varrho'   \in {\cal M}_d$.
\end{enumerate}
 \noindent\underline{\textbf{Output:}} The fraction of the volume of  ${\cal
M}_d$ describing the set of all  states in the constructed sign
permutation polytope possessing an entanglement type comparable to
the state $\rho \in {\cal M}_d$.

 \noindent\underline{\textbf{Procedure:}}
\begin{enumerate}
        \item  Let  $\tilde{\varrho'}_{} \in \E^{d^2-1}$ denote the Euclidean vector
    associated with an arbitrary  state $\varrho'_{} \in {\cal M}_d$.

        \item Let $T_{\tilde{\varrho}}(\tilde{\varrho'}_{})$
        be the image of $\tilde{\varrho'}_{}$
        under the mapping $T_{\tilde{\varrho}}(a): a - \tilde{\varrho}$,  $a \in
        \E^{d^2-1}$.
        \item Consider the positive cone  $\E_{{+}}^{d^2-1} := \{(\alpha_1,\dots,\alpha_{{d^2-1}}) \in \E^{d^2-1} \vert   \alpha_i \geq 0\}$. For suitable sign changes, let  $T_{+{\tilde{\varrho}}}(\tilde{\varrho'}_{})$ be the image of $T_{\tilde{\varrho}}(\tilde{\varrho'}_{})$ in the positive cone $\E_{{+}}^{d^2-1}$.
        \item Implement corollary \ref{thm5}: Letting ${\cal V}_{\pm\pi(a)}$ denote  the cross-polytope conv$\{\pi(\pm \alpha, 0, \dots, 0),$ $\pi \in S_{d^2-1}\}$
     of edge length $\sqrt{2}\alpha$.   Then
        \begin{eqnarray} T_{+ \tilde{\varrho}}(\tilde{\varrho'}_{}) \in
        {\cal V}_{\pm\pi(\alpha)} \Leftrightarrow T_{+\tilde{\varrho}}(\tilde{\varrho'}_{})
        \prec_w \alpha.\end{eqnarray}

\item If  $T_{+\tilde{\varrho}}(\tilde{\varrho'}_{}) \prec_w
\alpha$ then $T_{+\tilde{\varrho}}(\tilde{\varrho'}_{})$ is
contained in sign permutation polytope ${\cal
V}_{\pm\pi(\alpha)}$. Consequently,
$T_{+\tilde{\varrho}}(\tilde{\varrho'}_{})$ represents a state in
the
\begin{eqnarray}
\frac{2^{(2d+3)(d-1)/2}\alpha^{d^2-1}}{\sqrt{d}\pi^{d(d-1)/2}\Gamma(1)\dots\Gamma(d)}
\end{eqnarray}
fraction of all states contained in sign permutation polytope $
{\cal V}_{\pm\pi(\alpha)}$ which  possesses an entanglement type
comparable to  state $\rho \in {\cal M}_d$.
\end{enumerate}
\end{alg}
\begin{figure}
\begin{picture}(10,150)(0,0)
\put(100,){\includegraphics[width=6cm]{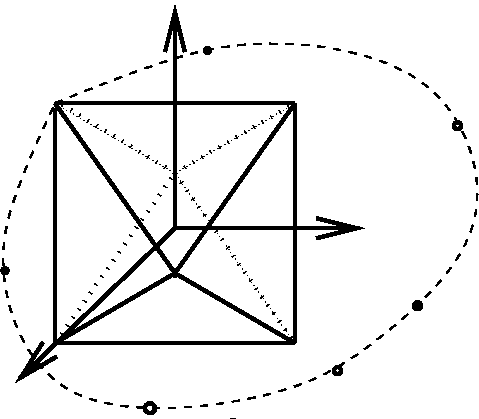}}
\put(255,145){${}_{{\E}^{d^2-1}}$}
\put(161,67.85){${}_{\star}$}
\put(167,140){${}_{T_{\tilde{\varrho}}(\tilde{\varrho}_1)}$}
\put(266,112){${}_{T_{\tilde{\varrho}}(\tilde{\varrho}_2)}$}
\put(276,107){${}_{.}$} \put(277,105){${}_{.}$}
\put(278,103){${}_{.}$}
\put(65,55){${}_{T_{\tilde{\varrho}}(\tilde{\varrho}_m)}$}
\put(150,106){${}_{T_{\tilde{\varrho}}(\tilde{\varrho'}_{})}$}
\put(150,96){$\ast$}
 \put(213,52){${\cal V}_{\pm\pi(a)}$}
\end{picture}
\caption{Translate of an arbitrary state ${\varrho'}_{}$,
$T_{\tilde{\varrho}}(\tilde{\varrho'}_{})$, contained in the
convex hull of the $(d^2-1)$-cross-polytope ${\cal
V}_{\pm\pi(a)}$, $\pi \in S_{d^2-1}$ with basis vector   $a =
(\alpha,0,\dots,0)$.} \label{figures2}
\end{figure}
Step (i) of algorithm \ref{algorithm2} considers a given arbitrary
state $\varrho'_{}\in {\cal M}_d$  to describe the set of states
differing from  $\varrho \in {\cal M}_d$. By analyzing the
classical representation of $\varrho' \in {\cal M}_d$, a
characterization of quantum states possessing an entanglement type
comparable to ${\cal V}_{\pm\pi(\alpha)}$ can be achieved. Step
(i) begins by evaluating the classical representation of
$\varrho'_{}$;  $\tilde{\varrho'}_{} \in \E^{d^2-1}$.  Step (ii)
implements the translation transformation $T_{\tilde{\varrho}}(a)$
on $\tilde{\varrho'}_{} \in \E^{d^2-1}$ thus translating
$\tilde{\varrho'}_{} \in \E^{d^2-1}$ to a neighbourhood of the
cross-polytope ${\cal V}_{\pm\pi(\alpha)} = {\rm conv}\{\pi(\pm
\alpha, 0, \dots, 0),$ $\pi \in S_{d^2-1}\}$. Step (iii) prepares
the point $\tilde{\varrho'}_{} \in \E^{d^2-1}$ for the  weak
majorization  criterion of corollary \ref{thm5} by placing
$\tilde{\varrho'}_{} \in \E^{d^2-1}$ in the positive cone
$\E_{{+}}^{d^2-1}$ for suitable sign changes. Step (iv) then
implements the weak majorization criterion for point inclusion in
a sign permutation polytope. If
$T_{+\tilde{\varrho}}(\tilde{\varrho'}_{}) \prec_w \alpha$ then
noting $T_{\tilde{\varrho}}(\tilde{\varrho'}_{})$ $\prec_w
T_{+\tilde{\varrho}}(\tilde{\varrho'}_{})$, it follows by  the
transitive property of weak majorization that
$T_{\tilde{\varrho}}(\tilde{\varrho'}_{}) \prec_w \alpha$, see
figure \ref{figures2}. Consequently,
$T_{\tilde{\varrho}}(\tilde{\varrho'}_{})$ $\in {\cal
V}_{\pm\pi(a)}$. Finally, step (v) returns the fraction of states
of ${\cal{M}}_d$ contained in the quantum cross-polytope ${\cal
V}_{\pm\pi(\alpha)}$ centered on the  state $\varrho \in {\cal
M}_d$. This fraction is given as ratio of the volume of the
quantum cross-polytope ${\cal V}_{\pm\pi(\alpha)}$ to the volume
of the set of all density matrices (\.{Z}yczkowski and Sommers
(2003)). \hfill $\Box$

Algorithm \ref{algorithm2} introduced the fraction of all states
of ${\cal{M}}_d$ contained in a quantum sign permutation polytope
possessing  an entanglement type equivalent to the entanglement
type of the state centered at the given quantum polytope.   This
fraction  was established in terms of the ratio of Hilbert-Schmidt
volumes. However, a Euclidean-$\epsilon$-ball argument describing
the distance between a given  state and a set of  states of
comparable entanglement type may also be considered. By lemma
\ref{lemma2}, the insphere of the constructed quantum  polytope
${\rm conv}\{\pi(\pm \alpha, 0, \dots, 0),  \pi \in S_{d^2-1}\}$
given in  algorithm \ref{algorithm1} is of radius
$\alpha/\sqrt{d^2-1}$. Letting $\epsilon = \alpha/\sqrt{d^2-1}$,
the set of  states $\varrho'_{} \in {\cal M}_d$ contained in
$\epsilon$-ball possessing an entanglement type comparable to that
of the
 state $\varrho \in {\cal M}_d$   is given by the
Hilbert-Schmidt distance
\begin{eqnarray}
\|\varrho'_{} -  \varrho\|_{\rm HS} \leq \alpha/\sqrt{d^2-1}.
\end{eqnarray}
Now we note  that the volume of an $n$-dimensional sphere of
radius $r$ is $V_n = {S_{n-1}r^n}/{n}$ with $S_{n-1} =
{2\pi^{n/2}}/{\Gamma(\frac{n}{2})}$ denoting the hyper-surface
area of an $n$-sphere of unit radius, and $\Gamma(\cdot)$ denoting
the Gamma function (Bengtsson and  \.{Z}yczkowski  (2006)). For
radius $r_{in} = \alpha/\sqrt{d^2-1}$,  the $(d^2-1)$-dimensional
insphere of the cross-polytope of edge length $\sqrt{2}\alpha$ has
volume
\begin{eqnarray}
 V_{d^2-1} = \frac{2\pi^{(d^2-1)/2}\alpha^{d^2-1}}{\Gamma(\frac{d^2-1}{2})(d^2-1)^{d^2+1}}.
\end{eqnarray}
On the other hand,  the volume of the $(d^2-1)$-dimensional cross-polytope of edge length $\sqrt{2}\alpha$ is ${(2\alpha)^{d^2-1}}/{(d^2-1)!}$. Comparing these volumes 
yields an estimate that the $(d^2-1)$-dimensional insphere of
radius $\alpha/\sqrt{d^2-1}$ is approximately
$(\pi/4)^{(d^2-1)/2}$ that of its respective cross-polytope.
Consequently, while the characterization of arbitrary states with
respect to polytope construction herein which possess an
entanglement type comparable to the given state $\varrho \in {\cal
M}_d$ is readily obtainable by both weak majorization and
Euclidean norm calculations,  arguing robustness of identification
in terms of  volume ratios between quantum sign permutation
polytopes and the space of density matrices may be of some merit.

Finally, the lower bound on the fraction of states introduced in
algorithm \ref{algorithm2}  may be improved by considering the
construction of the sign permutation polytope ${\cal{V}}_{\pm
\pi({\alpha})}$ with  $a =$
$(\alpha_1,\dots,\alpha_m,0_1,\dots,0_{d^2-1-m})$, $m \leq d^2-1$,
denoting a multiset of $k$ distinct elements of multiplicity $m_i\
(1\leq i \leq k)$. This approach is  more  costly and requires
that  ${2^mn!}/{({m_1!\dots m_{k-1}!(n-m)!})}$  vertices  be
evaluated as convex combinations of an initially given  convex set
of density matrices with a fixed entanglement type. Furthermore,
the resulting polytope structure  loses regularity exhibited by
the $n$-cube and $n$-cross-polytope thereby making volume
calculation difficult.

\section{Conclusion} \label{conclusion}
We  discussed the construction of   $n$-dimensional polytopes of
degree $n$ in the $n$-dimensional Euclidean space $\E^n$  called
sign permutation polytopes, and presented a  necessary and
sufficient condition that characterizes
 the  points of these polytopes.  We related the construction of  sign permutation polytopes in Euclidean space to the space of density matrices. As an application of such quantum polytopes, we considered the robustness of identifying   those states  of a fixed entanglement type. 
This process  proceeded in two stages, the first stage concerned
the construction of a convex polytope with a  vertex set of a
fixed type entanglement. The second stage  characterized via weak
majorization the fraction of  all quantum states  contained  in
the convex hull of a constructed convex polytope  which possess
an entanglement type comparable to the polytope's center state.

\ack The authors wish to thank  David Gross and  Eshan Momtahan
for helpful comments and suggestions. We gratefully thank DFG for
financial support.

\appendix
\section*{Appendix A}
\setcounter{section}{1}

\noindent{\bf{Proof of lemma \ref{lemma3.3}.}}
 Let us suppose otherwise. Then for ${x} \in
{\mathcal{V}}_{a}$ and ${y} \in {\mathcal{V}}_{b}$ there is a
hyperplane $H_{{u},\alpha}$ such that
\begin{eqnarray}
\langle {x},{u} \rangle  < \alpha \leq \langle {y}, {u}\rangle.
 \end{eqnarray}
Since $\sum_{k\leq n}{y_k} = \sum_{k\leq n}{\beta_k}$, we may take
the hyperplane $H_{1_n, \sum_{k\leq n}{y_k}}$, $1_n =
(1,\dots,1)\in \E^n$, to be the representative for the set of
hyperplanes $H_{{u}, \alpha}$ that place
 ${\mathcal{V}}_{b}$ in one of the closed halfspaces of $\E^{n}$, see Appendix B. In particular, ${\mathcal{V}}_{b}$ is
in the halfspace   $H^{+}_{1_n, \sum_{k\leq n}{y_k}}$ that satisfies $\sum_{k\leq n}{y_k} \leq \langle y, 1_n\rangle$.   
Now, $\langle {x}, 1_n \rangle = \sum_{k\leq n}{x_k}$. Consider
the sum $\sum_{k\leq n}{x_k}$.  For ${x} \in {\mathcal{V}}_{a}$
there are non-negative scalars $\lambda_i, i = 1, \dots, m$ with
$m\leq n!$ such that  $\sum_{i\leq m}\lambda_i = 1$ and ${x} =
 \sum_{i \leq m}{\lambda_i{a}_i}$. Therefore, we have it that
\begin{eqnarray}
 \sum_{k\leq n}{{x_k}} = \sum_{k\leq n}\sum_{i\leq m} {}\lambda_i \alpha_{ik} = \sum_{i\leq m}{}\lambda_i\sum_{k\leq n} {}\alpha_{ik}
= \sum_{i\leq m}{}\lambda_i\sum_{k\leq n} {} \ a_{k}, \label{sum}
\end{eqnarray}
since $\sum_{k\leq n} {}\alpha_{ik} = \sum_{k\leq n} {}\alpha_{k}$
for $i = 1, \dots, m$. Furthermore, since $\sum_{k\leq n}
{}\alpha_{k} = \sum_{k\leq n} {}\beta_{k}$, we have ${\alpha_j} =
\sum{}_{l \leq n}\beta_l - \sum{}_{\stackrel{l \leq n}{{l \ne
j}}}{\alpha_l}$ for $j \in \{1,\dots,n\}.$ It follows that
equation (\ref{sum}) may be written as
\begin{eqnarray}\label{simple} 
  \sum_{i\leq m}{}\lambda_i {}\left(\sum{}_{\stackrel{k \leq n}{{k \ne j}}}{\alpha_k}+\sum_{l
\leq n}{\beta_l} -  \sum{}_{\stackrel{l \leq n}{{l \ne
j}}}{\alpha_l} \right),\end{eqnarray}
 for $j \in \{1,\dots,d\}$. Simplifying,  we write equation (\ref{simple}) as $\sum_{i\leq m}{}\lambda_i\sum_{l\leq n} {}\beta_{l}$.
Since there are numbers $\mu_i \geq 0$ satisfying $\sum_{i\leq
m'}{\mu_i} = 1,$ we then have it that $\sum_{i\leq
m}{}\lambda_i\sum_{l\leq n} {}  \beta_{l}  =  \sum_{i\leq
m'}{}\mu_i\sum_{l\leq n} {}  \beta_{l}$. Moreover, since $\sum_{k
\leq n}{\beta_k} = \sum_{k \leq n}{\beta_{ik}}$, for $i =
1,\dots,m'$, it follows that $\sum_{i\leq m'}{}\mu_i\sum_{l\leq n}
{} \beta_{l}$ may be given by
\begin{eqnarray}
\sum_{i\leq m'}{}\mu_i\sum_{l\leq n} {}  \beta_{l} = \sum_{l\leq
n} {}\sum_{i\leq m'}{} \mu_i \beta_{l} = \sum_{l\leq n}
{}\sum_{i\leq m'}{} \mu_i \beta_{il} = \sum_{l\leq n} {y_l}.
\end{eqnarray}
Therefore, $\langle {x},1_n \rangle = \langle {y},1_n \rangle$,
and this contradicts the initial assumption that  the convex hulls
${\mathcal{V}}_{a}$ and ${\mathcal{V}}_{b}$ can be properly
separated. This completes the proof. \hfill$\Box$

\appendix
\section*{Appendix B}
\setcounter{section}{2}

  We construct the hyperplane $H_{{ u},\alpha}$ with the property $H_{{ u},\alpha} \cap {\mathcal{V}}_{ a} \ne \emptyset$.\\

\begin{figure}\begin{picture}(0,150)(0,0)
\put(70,-125){\includegraphics[width=8cm]{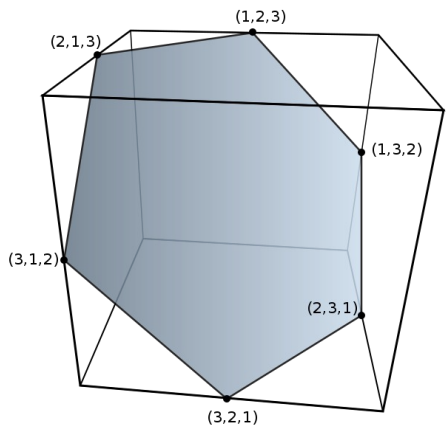}}
\end{picture}
\caption{Permutahedron: ${\mathcal{V}}_{a}$, $a = (1,2,3)$}
\label{va}
\end{figure}

\noindent {Example 1:} Constructing hyperplanes for the
permutahedron ${\mathcal{V}}_{\rm a}$ with  $\rm a = (1,2,3)$, see
figure \ref{va}.

Taking the points $(2,1,3)$ and $(2,3,1)$, we have it that the
parametric equation of hyperplane (line) containing these points
is given by $H_1 : (0,2,-2)s + (2,3,1)$, for $s \in \E$. For any
${ z} \in H_1$, we may write ${\rm z}$ componentwise as $z_1 = 2,
z_2 = 2s + 3, z_3 = -2s + 1$, $s \in \E$.  Now since a hyperplane
may be written in the form
\begin{eqnarray}
H_{{ u}, \alpha} = \left\{ { z} \in \E^d \  \vert \  \langle {
z},{ u} \rangle = \alpha \right\}
\end{eqnarray}
for ${\rm u} \in \E^d$\textbackslash$\{0\}$, and $\alpha \in
\R^d$, we may rewrite $H_1$ in terms of the  $H_{{u}, \alpha}$
form. We put $\alpha = \sum{a_k} = 6$.  Then $2u_1 + (2s+3)u_2 +
(-2s+1)u_3 = r$. In particular, we have it that $H_{{ u},6}$ is
constrained by $2u_1+4u_2 = 6$ with $u_2 = u_3$.  We determine the
remaining hyperplanes in a similar fashion. Proceeding
anticlockwise, $H_2 = (1,-1,0)s + (3,2,1)$, for $s \in \E$.
Rewriting $H_2$ in terms of an  $H_{{\rm u},6}$ expression, the
candidate vector ${\rm u} \in \E^3$ is constrained by the equation
$5u_1+u_3 = 6$ with $u_1 = u_2$. The hyperplane  $H_3 = (0,-1,1)s
+ (3,1,2)$, for $s \in \E$, rewriting $H_3$, the candidate vector
${ u} \in \E^3$ is constrained by the equation $3u_1+3u_2 = 6$
with $u_2 = u_3$.  The hyperplane  $H_4 = (-2,2,0)s + (1,3,2)$,
for $s \in \E$, rewriting $H_4$, the candidate vector ${ u} \in
\E^3$ is constrained by the equation $4u_1+2u_3 = 6$ with $u_1 =
u_2$. The hyperplane $H_5 = (0,1,-1)s + (1,3,2)$, for $s \in \E$,
rewriting $H_5$, the candidate vector ${ u} \in \E^3$ is
constrained by the equation $u_1+5u_2 = 6$ with $u_2 = u_3$.
Finally, the hyperplane  $H_6 = (-1,1,0)s + (1,2,3)$, for $s \in
\E$ and  rewriting $H_6$ in the form $H_{{ u},6}$, the candidate
vector ${ u} \in \E^3$ is constrained by the equation $3u_1+3u_3 =
6$ with $u_1 = u_2$. Note that while each evaluated hyperplane has
infinitely many solutions, a candidate that satisfies each of the
particular constraints is the  vector with components all equal to
1. Of course, since  the direction vectors in each of the
parametric expressions are permutations of each other then a
candidate vector ${u} \in \E^3$ which satisfies all of the above
constraints is precisely that vector which, under permutation of
indices corresponding to the direction vectors, generates a
solution vector  ${u'} \in \E^3$ for all hyperplane
$H_{{ u'},6}$.  The only such candidate is the vector $u=(1,1,1)$.\\

\noindent {Example 2:} {Constructing hyperplanes for
${\mathcal{V}}_{ a}$, ${ a} = (1,2,\dots,n)$}.

 Continuing in the manner of above, we seek a candidate
hyperplane for the set of all hyperplanes bounding
${\mathcal{V}}_{ a}$ entirely in a closed halfspace. We claim the
hyperplane given by $H_{1_n,\frac{n(n+1)}{2}}$, $1_n =
(1,\dots,1)\in\E^n$, is such a candidate. Taking the vector ${a} =
$ $(1,2,\dots,n)$ $ \in \E^n$, the associated permutahedron
${\mathcal{V}}_{ a}$ is defined as the convex hull of all vertices
obtained by permuting the coordinates of the vector ${a}$. Since
the set of vertices connected by and edge to the vector ${a}$ are
exactly those permutations of vector ${ a}$ that differ by an
adjacent transposition, we have it that the set of vertices
connected by an edge  to vertex $a$ is  the set
$\{(1,2,\dots,n-2,n,n-1)$, $(1,2,\dots,n-1,n-2,n)$, $\dots,
(2,1,\dots,n)\}$. Taking the vertex $ a \in \E^n$ together with
any $n-2$ vertices of the connecting set, of which there are $n-1$
possible choices, determine a $(n-2)$-dimensional hyperplane. Fix
the hyperplane $H_1$ to contain the set of vertices connected by
an edge to the vector $a \in \E^n$ that exclude the last vertex
$(2,1,3,\dots,n)$. The parametric equation for the hyperplane
$H_1$ is written in terms of the directions vectors
$\{(0,\dots,0,1,-1), (0,\dots,0,1,-1,0), \dots,
(0,1,-1,0\dots,0)\}$. Each direction vector is represented as the
difference between the vertex $a$ and an element of the set
$\{(1,2,\dots,n-2,n,n-1)$, $(1,2,\dots,n-1,n-2,n)$, $\dots,
(1,3,2,\dots,n)\}$. Rewriting $H_1$ in the $H_{{ u},\alpha}$ form
with $\alpha = \frac{n(n+1)}{2}$, we have it that the candidate
vector ${ u} \in \E^n$ is constrained by the equation $u_1 +
\left(\frac{n(n+1)}{2} - 1\right)u_2 = \frac{n(n+1)}{2}$ with $u_2
= u_3 = \dots = u_n$. Next, the hyperplane $H_2$ determined by the
vertex ${ a} \in \E^n$ and its set of connecting vertices that
excludes the vertex $(1,3,2,4,\dots,n)$ has its parametric
expression written as linear combination of the set of direction
vectors $\{(0,\dots,0,1,-1), (0,\dots,0,1,-1,0), \dots,
(0,0,1,-1,0\dots,0)$, $(1,-1,0,\dots,0)\}$. Writing  the
parametric equation of the hyperplane  $H_2$ in terms of the $H_{{
u}, \frac{n(n+1)}{2}}$ form, we have the candidate vector ${ u}
\in \E^n$ constrained by $u_1 + \left(\frac{n(n+1)}{2}-1\right)u_3
= \frac{n(n+1)}{2}$ with $u_1 = u_2$ and $u_3 = u_4 = \dots =
u_n$. In a similar manner, the remaining $n-3$ hyperplanes
containing the vertex $ a$ can be formed. Solving for the  vector
${ u} \in \E^n$ that determines the hyperplane $H_{ u,
\frac{n(n+1)}{2} }$ associated with $H_1$ and $H_2$ yields the
 vector $1_n = (1,\dots,1)\in \E^n$. It follows that $H_{1_n, \frac{n(n+1)}{2} }$  may be
taken as the representative for the set of hyperplanes containing
vertex $a \in \E^n$. Furthermore, since all vertices $ a' \in
\E^n$ of the permutahedron ${\mathcal{V}}_{a} \subset \E^n$ are
permutations of the vertex $ a$, for $\pi_{ a'} \in
{\mathcal{S}}_n$, then the set of hyperplanes containing the point
$ a' \in \E^n$ can be rewritten in terms of a hyperplane $H_{ u,
\frac{n(n+1)}{2} }$ for some candidate vector $u \in \E^d$. For
such a candidate $ u \in \E^n$ and permutation $\pi^{-1}_{ a'} \in
{\mathcal{S}}_n$, $H_{\pi^{-1}_{ a'}({ u}), \frac{n(n+1)}{2}}$ is
a representative hyperplane for the set of hyperplane containing
the vertex $ a \in \E^n$. Consequently, $H_{\pi^{-1}_{ a'}({ u}),
\frac{n(n+1)}{2}} = H_{1_n,\frac{n(n+1)}{2}}$. In particular,
$\pi^{-1}_{ a'}({ u}) = 1$.  Since the vector with all components
equal to 1 is invariant under permutations of the indices, it
follows that $H_{1,\frac{n(n+1)}{2}}$ may be taken as a
representative for the set of all hyperplanes bounding
${\mathcal{V}}_{ a}$ entirely in a closed halfspace with the
property that $H_{z,\alpha} \cap {\mathcal{V}}_{ a} \ne
\emptyset$.

\section*{References}
\begin{harvard}

\item[] Ac\'{i}n A, Andrianov A, Jane E and Tarrach R 2001
Three-qubit pure-state canonical forms \emph{J. Phys. A: Math.
Gen.} \textbf{34}  6725-39

\item[] Bishop E and  Bridges D 1985  \emph{Constructive Analysis}
(Berlin: Springer) (1985)

 \item[] Bengtsson I and \.{Z}yczkowski K 2006 \emph{Geometry of quantum states} (New York: Cambridge University Press)

\item[] Bennett C H, Bernstein H J, Popescu S and Schumacher B
1996 {Concentrating partial entanglement by local operations}
\emph{Phys. Rev.} A \textbf{53}  2046-52


\item[] Br\o{}ndsted 1983 {An introduction of convex polytopes}
(New York: Springer-Verlag)

\item[] Cayley A 1845  {On the theory of linear transformations}
\emph{Cambridge Math. J.} \textbf{4} 193-209

\item[] Coffman V, Kundu J and Wootters W K 2000 {Multipartite
pure-state entanglement and the generalized GHZ states}
\emph{Phys. Rev.} A  \textbf{61} 052306

\item[] Coxeter H S M  1969 {Introduction to Geometry} 2nd ed.
(New York: John Wiley \& Sons, Inc.)

\item[] Coxeter H S M  1948 {Regular Polytopes} (London: Methuen
\& Co. Ltd.)


\item[] D\"ur W, Vidal G and Cirac J I 2000 {Three qubits can be
entangled in two inequivalent ways} \emph{Phys. Rev.} A
\textbf{62} 062314


\item[] Gaiha P and  Gupta S K 1977 {Adjacent vertices on a
permutohedron} \emph{SIAM J. Appl. Math.} \textbf{32} 323-327



\item[] Gr\"unbaum 2003 {Convex polytopes} 2nd ed. (New York:
Springer-Verlag)


\item[] Hardy G H, Littlewood J E and P\'{o}lya G 1934
{Inequalities} (New York: Cambridge University Press)

\item[] Kampermann K, G\"uhne O, Wilmott C and Bruss D 2010
{unpublished}


\item[] Markus 1964 {Eigenvalues and singular values of sum and
product of linear operators} \emph{Russian Math. Surveys}
\textbf{19}  91-120

\item[] Marshall A W and Olkin I 1979 {Inequalities: theory of
majorization and its applications} (New York: Academic Press)

\item[] Matousek J 2002 {Lectures on Discrete Geometry} (New York:
Springer-Verlag)

\item[] Mirsky L 1959 {On a convex set of matrices} \emph{Arch.
Math.}  \textbf{10} 88-92


 \item[] O' Rourke J 1998 {Computational Geometry in C} (New York: Cambridge University Press)

\item[] Vidal G, D\"ur W and Cirac J I 2000 {Reversible
combination of inequivalent kinds of multipartite entanglement}
\emph{Phys. Rev. Lett.} \textbf{85} 658-61


\item[] Rado R 1952 {An equality} \emph{J. London Math. Soc.}
\textbf{27}  1-6

\item[] Ziegler G M 1995 {Lectures on polytopes} (Berlin:
Springer-Verlag)

 \item[] \.{Z}yczkowski K  and Sommers H -J  2003 {Hilbert schmidt volume of the set of mixed quantum states} \emph{J. Phys. A} \textbf{36}  10115-30

\end{harvard}

\end{document}